\documentclass[10pt, conference, letterpaper]{IEEEtran}
\def\BibTeX{{\rm B\kern-.05em{\sc i\kern-.025em b}\kern-.08em
    T\kern-.1667em\lower.7ex\hbox{E}\kern-.125emX}}
\IEEEoverridecommandlockouts
\usepackage{algorithm,amsbsy,amsmath,amssymb,epsfig,bbm,mathrsfs,fancyhdr,fancyvrb,url,color,soul}
\usepackage{amsthm}
\usepackage{graphicx}
\usepackage{cite}
\usepackage{float}
\usepackage{mathtools}
\usepackage{algorithm}
\usepackage{algpseudocode}
\usepackage{epstopdf}
\usepackage{epsfig}
\usepackage{tabularx}
\usepackage{pgfplots}
\usepackage{subcaption}
\usepackage{bm}
\usepackage{multicol}
\usepackage{makecell}
\usepackage{afterpage}

\usepackage{caption}
\usepackage[font=small]{caption} 
\usepackage{array}

\usepackage[english]{babel}
\newcolumntype{M}[1]{>{\centering\arraybackslash}m{#1}}

\title{Performance Analysis of OTFS-NOMA System with Fractional Doppler}
\author{\IEEEauthorblockN{Wafa Hedhly, Leila Musavian, and Nikolaos Thomos}
 \thanks{The authors are with the School of Computer Science and Electronic Engineering, University of Essex, Wivenhoe Park, Colchester CO43SQ, United Kingdom (e-mail: { \{wafa.hedhly, leila.musavian, nthomos\}@essex.ac.uk).} This work was funded by the Engineering and Physical Science Research Council, UK, [grants number EP/X012204/1, EP/X04047X/2, EP/Y037243/1].}
 }
\begin{document}

\maketitle

\begin{abstract}
    In this work, we investigate the effect of fractional Doppler on the performance of a system using orthogonal time frequency space (OTFS) modulation and non-orthogonal multiple access (NOMA) where users have different mobility profiles. Fractional Doppler results in inter-Doppler interference (IDI) and degrades the quality of OTFS-modulated signals.
    We consider a downlink (DL) communication scenario where multiple users are distinguished based on their mobility profiles into a single high-mobility (HM) user and multiple low-mobility (LM) users. OTFS modulation is implemented for the HM user by embedding its information symbols in the delay-Doppler domain, while LM users' symbols are represented in the time-frequency (TF) domain.
    The LM users' signals are kept orthogonal to each other in the frequency domain by accessing disjoint subcarriers.
    Further, NOMA spectrum sharing is implemented between the HM user and the KM users to achieve higher spectral efficiency.
    Performance analysis in terms of DL spectral efficiency and outage probability is conducted for different system parameters.
    The numerical results show that IDI has a noticeable performance impact on the HM user, depending on the NOMA parameters. 
  
\end{abstract}
\section{Introduction}
Transportation systems are experiencing unprecedented progress in terms of speed and connectivity, where high-speed rail systems are traveling at tremendous speeds, and vehicles are migrating towards autonomy. Consequently, modern vehicular networks are expected to accommodate the ever-increasing demands of high-mobility users and deal with users having diverse mobility profiles while simultaneously addressing the rising communication challenges, mainly fast channel variation and high Doppler spreads. 
In this regard, OTFS emerged as a novel modulation scheme that transforms the fast-varying time-frequency (TF) representation of the wireless channel to a quasi-static sparse alternative in the delay-Doppler (DD) domain. This representation captures the geometry of the channel and outperforms orthogonal frequency division multiplexing (OFDM) in high-mobility scenarios.
As far as multiple access is concerned, the non-orthogonal multiple access (NOMA) paradigm proved its benefits in multiuser networks with increased spectral efficiency compared to its orthogonal counterparts.  NOMA is particularly beneficial in high-mobility scenarios since it allows spreading the OTFS frame across the whole time-frequency resource block and thus improving OTFS resolution and detection.
This renders a combination of OTFS modulation and NOMA scheme very beneficial for vehicular networks with users having heterogeneous mobility profiles. Nevertheless, when representing the channel taps in the DD domain, the Doppler tap does not necessarily fall on an integer index, resulting in a fractional Doppler \cite{raviteja2018interference}. 
This results in inter-Doppler interference. Hence, evaluating this additional source of interference is necessary to reap the maximum benefit of OTFS-NOMA systems.

OTFS-NOMA systems have recently become the focus of multiple research efforts \cite{ding2019otfs,ding2019robust,ge2021otfs,zhou2022active,zhou2021joint,hu2024cross,mcwade2023low,hedhly2024otfs}. In \cite{ding2019otfs}, Ding et al. proposed an OTFS-NOMA system that leverages the benefits of both OTFS modulation and NOMA considering a single HM user and multiple LM users. MIMO beamforming was optimized in \cite{ding2019robust} in a similar to \cite{ding2019otfs} OTFS-NOMA setup.
Uplink transmission was considered in \cite{ge2021otfs} where OTFS modulation was used for all users in conjunction with NOMA. OTFS-NOMA was adopted in \cite{zhou2022active} and \cite{zhou2021joint} to provide access to a massive number of IoT devices and mitigate the high Doppler spread of satellite-to-ground communications. The authors of \cite{hu2024cross} utilized cross-domain channel estimation to propose a novel approach for successive interference cancellation in OTFS-NOMA systems. In \cite{mcwade2023low}, a low-complexity user equalization and detection scheme was proposed for OTFS-NOMA systems to mitigate multi-user interference while considering fractional Doppler. However, the effect of this source of interference was not evaluated. The authors of \cite{hedhly2024otfs} considered an OTFS-NOMA system where a base station (BS) utilizes multiple input multiple output (MIMO) beamforming to communicate with HM and LM users grouped into clusters.

The aforementioned research works neglect the effect of fractional Doppler on the performance of OTFS-NOMA systems, which we study in this paper.
Specifically, in this work, we investigate the impact of inter-Doppler interference (IDI) resulting from fractional Doppler on the performance of OTFS-NOMA system with users having heterogeneous mobility profiles. To this end, we consider a multi-user communication network where a BS communicates with a single HM user and multiple LM users spread across a cell. The information symbols of the HM user are placed in the DD domain to combat the high Doppler spreads of its channel, while the information symbols of the LM users are placed in the TF domain. All signals are multiplexed in the same resource block using NOMA, but LM users' signals are orthogonal to each other in the frequency domain. The contributions of this paper can be summarized as follows:
\begin{itemize}
    \item We model the proposed OTFS-NOMA system while taking into account the fractional Doppler parameters in the received signal. We distinguish between the desired signal free from IDI and the IDI signal.
    \item We apply a minimum mean square error (MMSE) detector at each user's receiver in order to detect the desired signal of the HM user first. Then, after successfully decoding the HM user's signal, the LM user's signal is detected in the TF domain.
    \item We derive the closed-form expressions of detection signal-to-noise ratios (SNRs) with the additional IDI interference term at the HM side.
    \item We evaluate the performance of the system in terms of spectral efficiency and outage probability. We particularly assess the impact of the IDI and the NOMA power allocation factors in the presence of IDI on the performance of HM and LM users.
\end{itemize}

%\clearpage
\section{System Model}
  \begin{figure}[!h]
	\centering 
	\captionsetup{justification=centering,margin=1cm}
	\includegraphics[width=3.5in,trim={1cm 0 1cm 1cm},clip]{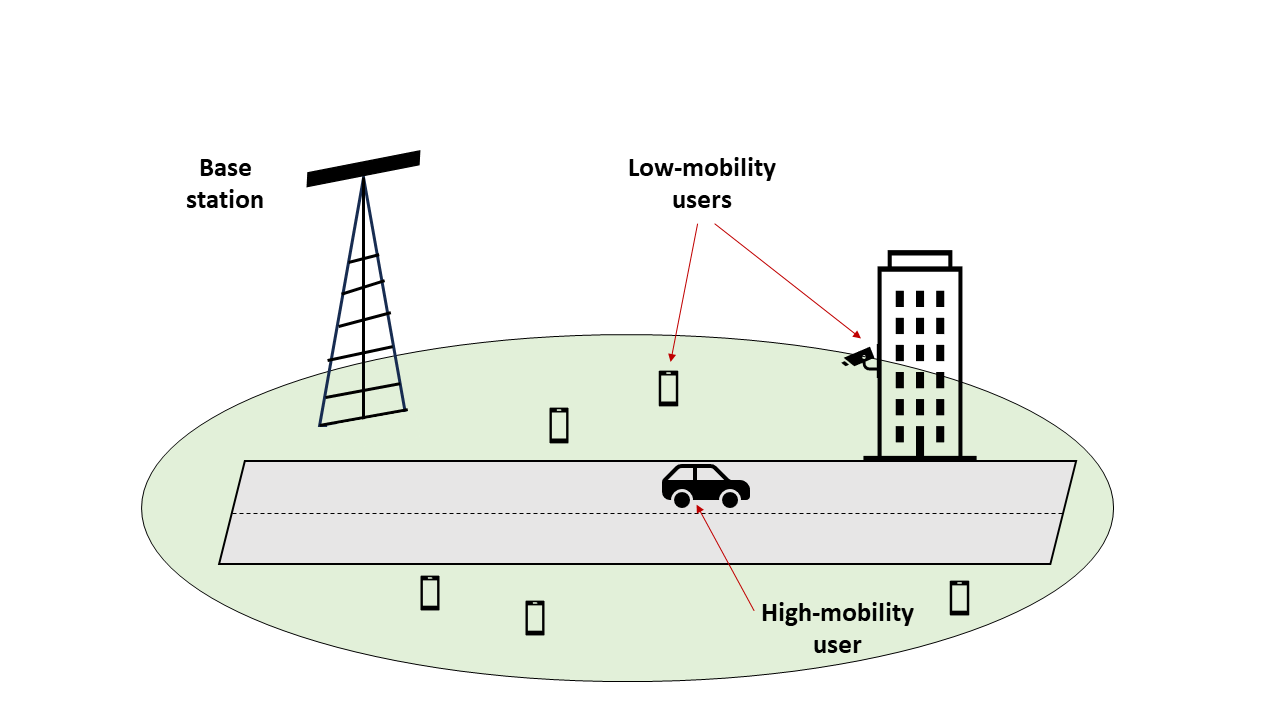}
	\caption{System Setup.}
	\label{system}
\end{figure}
In this work, we consider a DL scenario where a BS communicates with a single HM user and $U$ LM users. The BS is equipped with $A$ antennas, while all users are equipped with single antennas. The HM user's signal is modulated as an OTFS frame, while LM users' signals are OFDM modulated. The HM user's signal and all LM users'signals are multiplexed using NOMA paradigm where the power allocation factor of user $u$ is $p_u$ for $0 \le u \le U$ (where $u = 0$ indicates the HM user) and $\sum_{u = 0}^U p_u = 1$. LM users signals are orthogonal amongst each other by accessing disjoint subchannels similarly to \cite{ding2019otfs}.

An OTFS frame is transmitted to the HM user in the DD domain comprising $NM$ information symbols, with $M$ delay bins and $N$ Doppler bins. 
If we consider $T$ as the sampling time interval and $\Delta f$ as the spacing between subchannels, the duration and bandwidth of the transmitted frame are $NT$ and $M \Delta f$, respectively, where $\Delta f = 1/T$. 
The transmitted signal to the HM user in the DD domain $s_0^{\mathrm{DD}}$ can be converted to the TF domain as follows,
\begin{equation}
    s_{0}^{\mathrm{TF}}[n,m] = \frac{1}{NM} \sum_{k = 0}^{N - 1} \sum_{l = 0}^{M - 1} s_{0}[k,l] e^{j2\pi \left( \frac{k n}{N}-\frac{m l}{M} \right)},
\end{equation}
where $0 \le n \le N-1$ and $0 \le m \le M-1$ and $k$ and $l$ are the Doppler and delay bins, respectively.
The transmitted signal to each LM user $u = 1 \ldots U$ is expressed in the TF domain as follows,
\begin{align}\label{su}
    \mathbf{s}_{u}^{\mathrm{TF}} [n,m] = 
    \begin{cases}
        & \mathbf{s}_{u}^\mathrm{TF}[n, u-1], \quad \mathrm{if} \; m = u - 1, \\
        & 0, \quad \mathrm{otherwise},
    \end{cases}
\end{align}
where $0 \le n \le N - 1$ and $0 \le m \le M - 1$.
The users'signals are multiplexed using NOMA principle. Therefore, the aggregate transmitted signal from the BS is expressed in the DD domain as follows,
\begin{equation}
    {s}[k,l] = \sum_{u = 0}^{U} \sqrt{p_u} {s}_{u}^{\mathrm{DD}}[k,l],
\end{equation}
where ${s}_{u}^{\mathrm{DD}}$ is the baseband signal for user $u$ expressed in the DD domain.

The overall transmitted signal is expressed as follows,
\begin{equation}\label{xa}
    {x}[k,l] =  \sum_{a = 1}^A {v}_a {s}[k,l],
\end{equation}
where $v_a$ is a weighing factor for antenna element $a$.

In this work, we assume the presence of fractional Doppler, which indicates that the path tap does not fall on an exact Doppler bin. In this case, the Doppler and delay taps of path $p$ where $1 \le p \le L_0$ of the HM user channel are expressed as,
\begin{align}\label{taps}
    \nu_{p,0} &= \frac{k_{p,0} + \kappa_{p,0}}{NT}  \nonumber  \\ \tau_{p,0} &= \frac{l_{p,0}}{M \Delta f},
\end{align}
where $k_{p,0}$ and $l_{p,0}$ are the Doppler tap and delay tap of path $p$, $-1/2 < \kappa_{p,0} \le 1/2$ is the fractional Doppler of path $p$ expressed as a real number.
We assume $L_{u}$ denotes the number of paths of the wireless channel of user $u$. If we denote the number of paths of LM user $u$ as $L_u$, each path has a delay of $\tau_{p,u}$. LM users do not have Doppler shifts. 
%When fractional Doppler is taken into account, each scatterer in the medium results in multiple subpaths, including one dominant path.
%We assume that each path $p$ of the HM user channel has $S_p^q$ subpaths.

The received signal at the HM user from antenna element $a$, $1 \le a \le A$ can be written in the DD domain \cite{raviteja2018interference} as follows,
\begin{align}\label{y0a}
    {y}_{0,a}[k,l] &= \sum _{p=1}^{L_0} \sum _{q =-N_{p}}^{N_{p}}  h_{p,a,0} (q) \times \nonumber\\
    &{x}_a \left [{\left(k-k_{p,0}+q \right)_{N}, \left(l - l_{p,0}\right)_{M}}\right]\!+ w_{0}[k,l].
\end{align}
where $w_{0}$ denotes the complex Gaussian additive noise with zero-mean and variance $\sigma^2$ and $h_{p,a,0} (q) $ is the channel coefficient of each subpath $q, 0 \le q \le N_p$ expressed as follows,
\begin{equation}
    h_{p,a,0} (q) = \alpha_{p,a,0} e^{-j2\pi \nu_{p,0} \tau_{p,0} } \left ({\frac {e^{-j {2\pi } (-q - \kappa _{p,0}) }-1}{N e^{-j \frac {2\pi }{N} (- q - \kappa _{p,0})}-N}}\right),
\end{equation}
where $\alpha_{p,a,0} \sim \mathcal{CN}(0, 1/L_0)$ is the channel gain of path $p$ between antenna $a$ of the BS and the HM user.
Herein, we assume that the transmit and receive waveforms are orthogonal and that the delay and Doppler taps locations are the same for all antennas.

The fractional Doppler results in a channel representation where $2N_p + 1$ subpaths are generated from each main path $p$ where $N_p << N$. Among these subpaths, $2N_p$ signals are interfering signals to neighboring symbols in the OTFS grid. This results in IDI. We assume the number of subpaths is equal for all $1 \le p \le L_0$ denoted as $N_0$.\\
% \begin{figure*}[b]
% \begin{align}\label{y0a}
%     {y}_{0,a}[k,l] = \sum _{p=1}^{P_0} \sum _{q =-N_{p}}^{N_{p}}  \alpha_{p,a,0} e^{-j2\pi \nu_{p,0} \tau_{p,0} } \left ({\frac {e^{-j {2\pi } (-q - \kappa _{p,0}) }-1}{N e^{-j \frac {2\pi }{N} (- q - \kappa _{p,0})}-N}}\right)    {x}_a \left [{\left(k-k_{p,0}+q \right)_{N}, \left(l - l_{p,0}\right)_{M}}\right]\!+ w_{0}[k,l].\\
%     \noindent\rule{\textwidth}{1pt}
%     \nonumber
% \end{align}
% \end{figure*}
%The signal indexed by $q = 0$ is the signal not affected by fractional delay and has a major contribution.  
Remark: Increasing the OTFS frame duration $NT$ reduces the IDI since it minimizes the number of non-integer Doppler taps at the expense of increased signal latency \cite{wei2022orthogonal}.

The received signal in \eqref{y0a} can be decomposed into a desired signal and an IDI signal as follows,
\begin{equation}
   {y}_{0,a} = {y}_{\mathrm{M},a} + {y}_{\mathrm{I},a},
\end{equation}
where,
\begin{align}\label{M_IDI}
    &{y}_{\mathrm{M},a}[k,l] = \sum _{p=1}^{L_0}  h_{p,a,0} (0) {x}_a \left [{\left(k-k_{p,0} \right)_{N}, \left(l - l_{p,0} \right)_{M}}\right] \nonumber \\
    &{y}_{\mathrm{I},a}[k,l] = \sum _{p=1}^{L_0}\! \sum _{ \substack{q = -N_p \\ q\neq 0}}^{N_{p}} \! h_{p,a,0} (q)  {x}_a \! \left [{ \left(k-k_{p,0}+q\right)_{N} \!, \!\left(l - l_{p,0} \right)_{M}}\right].
\end{align}
Thus, the received signal at the HM user in a vector form can be written as,
\begin{equation}\label{y0}
    \mathbf{y}_0 = \sum_{a = 1}^A \mathbf{H}_{\mathrm{M},a} \mathbf{x}_a + \sum_{a = 1}^A \mathbf{H}_{\mathrm{I},a} \mathbf{x}_a  + \mathbf{w}_0,
\end{equation}
where $\mathbf{H}_{\mathrm{M},a}, \mathbf{H}_{\mathrm{I},a}  \in \mathcal{C}^{N\!M \times N\!M}$ are the corresponding block circulant ({$M$ circulant blocks of $N \times N$ circulant matrices}) channel matrices to the desired and IDI signals, respectively. The DD-domain noise in a vector form is $\mathbf{w}_{0} \sim \mathcal{CN}(\mathbf{0}, {\sigma}^2 \mathbf{I}_{N \! M})$.
%the main and interference block circulant channel matrices between user $u$ and antenna $a$, respectively, constructed from \eqref{M_IDI} ($M$ circulant blocks of $N \times N$ circulant matrices) and $\mathbf{w}_{0} \sim \mathcal{CN}(\mathbf{0}, {\sigma}_{\mathrm{w}}^2 \mathbf{I}_{N \! M})$ is the i.i.d additive noise vector in the DD domain. The $N \! M \times 1$ vector $ \mathbf{y}_{0}$ is constructed similarly to $\mathbf{x}_a$. 
The following condition has to be satisfied while constructing the vector $\mathbf{x}_{a}$ from the scalar form expressed in \eqref{xa}: the $(k+Nl)^{\mathrm{th}}$ element of $\mathbf{x}_{a}$ is equal to $x_a [k,l]$. The vectors $\mathbf{y}_{0}$ and $\mathbf{w}_{0}$ are similarly constructed.

The received signal in \eqref{y0} can be re-written equivalently as follows,
\begin{align}\label{y0q}
    \mathbf{y}_0 &= \sqrt{p_0} \sum_{a = 1}^A v_{a} \mathbf{H}_{\mathrm{M},a} \mathbf{s}_0 + {\sqrt{p_0} \sum_{a = 1}^A v_{a} \mathbf{H}_{\mathrm{I},a} \mathbf{s}_0}  \nonumber \\ 
    &+ {\sum_{a = 1}^A v_{a} \mathbf{H}_{a,0} 
 \sum_{i = 1}^{U} \sqrt{p_i} \mathbf{s}_{i}}  + \mathbf{w}_0,
\end{align}
where $\mathbf{H}_{a,0}$ is the overall channel matrix constructed from \eqref{y0a}. Note that the OTFS channel matrix in the presence of fractional Doppler is less sparse.
%Due to the fractional Doppler, the resulting OTFS matrix has less non-zero elements than in the case of the absence of the fractional delay since the number of non-zero elements per row and column of the channel matrix is $N_{\mathrm{z}} = \sum_{p = 1}^{P_0^q} (2N_p + 1)$. 

Since LM users are not subject to IDI, the received signal at user $u, \; 1 \le u \le U$ can be expressed as,
\begin{equation}\label{yu}
    \mathbf{y}_u  = \sqrt{p_0} \sum_{a = 1}^A v_a \mathbf{H}_{u,a} \mathbf{s}_0 + \sum_{a = 1}^A v_{a} \mathbf{H}_{a,u}  
 \sum_{i = 1}^{U} \sqrt{p_i} \mathbf{s}_{i}  + \mathbf{w}_u,
\end{equation}
where $\mathbf{H}_{u,a}$ is the DD-domain channel matrix between user $u$ and antenna $a$ and $\mathbf{w}_{u}\sim \mathcal{CN}(\mathbf{0}, {\sigma}^2 \mathbf{I}_{N \! M})$ is the noise vector.

\section{Signals Detection and Performance}
%\section{User Detection in the Presence of Fractional Doppler}
In this section, we present the detection of users using MMSE.
We consider the following matrix expressions,
\begin{align}\label{H0}
    \mathbf{H}_\mathrm{M} &= \sum_{a = 1}^A v_a\mathbf{H}_{\mathrm{M},a},  \nonumber  \\
    \mathbf{H}_u &= \sum_{a = 1}^A v_a \mathbf{H}_{a,u}.
\end{align}
\subsection{Detection at the HM User}
The HM receiver applies the following MMSE detector to detect $\mathbf{s}_0$,
\begin{equation}
    \mathbf{G}_0 =  \left( {\mathbf{H}}_\mathrm{M}^\mathrm{H} \mathbf{H}_\mathrm{M}  + \rho \mathbf{I}_{NM}\right)^{-1} {\mathbf{H}}_\mathrm{M}^\mathrm{H},
\end{equation}
where $\rho > 0$ and $\mathbf{I}_{NM}$ is an $NM \times NM$ identity matrix.
Using the diagonalization of block-circulant matrices as in \cite{singh2022ber}, the detection matrix $\mathbf{G}_0$ can be rewritten as,
\begin{equation}
    \mathbf{G}_0 = \mathbf{\Psi}^\mathrm{H} \left( {\mathbf{D}}_\mathrm{M}^{\mathrm{H}} \mathbf{D}_\mathrm{M} + \rho \mathbf{I}_{NM} \right)^{-1} {\mathbf{D}}_\mathrm{M}^{\mathrm{H}} \mathbf{\Psi},
\end{equation}
where $\mathbf{D}_{M}$ is a diagonal matrix expressed as,
\begin{equation}
    \mathbf{D}_\mathrm{M} =  \sum_{a = 1}^A v_a \mathbf{D}_{\mathrm{M},a},
\end{equation}
where $\mathbf{D}_{\mathrm{M},a}$ can be calculated from the following diagonalization method of block-circulant matrices as,
\begin{equation}
    \mathbf{H}_{\mathrm{M},a} =  \mathbf{\Psi}^\mathrm{H} \mathbf{D}_{\mathrm{M},a} \mathbf{\Psi},
\end{equation}
where $\mathbf{\Psi} = \mathbf{F}_\mathrm{N} \otimes \mathbf{F}_\mathrm{M}$ is a unitary matrix, $\mathbf{F}_\mathrm{N}$ and $\mathbf{F}_\mathrm{M}$ are discrete Fourier transform (DFT) matrices. 

If we denote by $\lambda_{a,i}^{\mathrm{M}}, 1 \le i \le NM$ the eigenvalues of $\mathbf{H}_{\mathrm{M},a}$, then the matrix $\mathbf{G}_0$ can be equivalently written as,
\begin{equation}
    \mathbf{G}_0 = \mathbf{\Psi}^{\mathrm{H}} \mathbf{\Delta}_0 \mathbf{\Psi},
\end{equation}
where $\mathbf{\Delta}_0$ is a diagonal matrix consisting of elements,
%end of plagiarism
\begin{equation}\label{delta0}
    \delta_{0,i} = \frac{\sum_{a = 1}^A   ({v_a} \lambda_{a,i}^{\mathrm{M}})^*}{ \left |\sum_{a = 1}^A  v_{a} \lambda_{a,i}^{\mathrm{M}} \right|^2 + \rho }, 1 \le i \le NM.
\end{equation}
Similarly, the matrices $\mathbf{H}_{\mathrm{I},a}$ and $\mathbf{H}_{a,0}$ can be decomposed as follows,
\begin{align}
    \mathbf{H}_{\mathrm{I},a} &= \mathbf{\Psi}^\mathrm{H} \mathbf{D}_{\mathrm{I},a} \mathbf{\Psi}, \nonumber \\
    \mathbf{H}_{a,0}  &= \mathbf{\Psi}^\mathrm{H} \mathbf{D}_{a,0} \mathbf{\Psi},
\end{align}
where $\mathbf{D}_{\mathrm{I},a}$ and $\mathbf{D}_{a,0}$ are diagonal matrices comprising the eigenvalues of $\mathbf{H}_{\mathrm{I},a}$ and $\mathbf{H}_{a,0}$ respectively.\\
%As a result, the proposed equalizer is applied to the received signal at the HM user in \eqref{y0q} as follows,
As a result, the received signal at the HM user \eqref{y0q} is detected using the proposed equalizing matrix as follows,
\begin{equation}\label{Gy}
    \mathbf{G}_0 \mathbf{y}_0 = \sqrt{p_0} \mathbf{E}_0 \mathbf{s}_0 +\sqrt{p_0}  \mathbf{F}_0 \mathbf{s}_0  + \mathbf{T}_0 \sum_{i = 1}^{U} \sqrt{p_i}  \mathbf{s}_i +  \mathbf{z}_0,
\end{equation}
where $\mathbf{z}_0 = \mathbf{G}_0 \mathbf{w}_0 $ and the matrices $\mathbf{E}_0$, $\mathbf{F}_0$ and $\mathbf{T}_0$ are written as follows,
%$ \mathbf{E}_0 = \mathbf{\Psi}^{\mathrm{H}} \mathbf{\Delta}_0 \mathbf{D}_0 \mathbf{\Psi}$, $ \mathbf{F}_0 = \mathbf{\Psi}^{\mathrm{H}} \mathbf{\Delta}_0 \sum_{a = 1}^A v_a \mathbf{D}_{\mathrm{I},a} \mathbf{\Psi}$ and $\mathbf{T}_0 = \mathbf{\Psi}^{\mathrm{H}} \mathbf{\Delta}_0 \sum_{a = 1}^A v_a  \mathbf{D}_{a,0} \mathbf{\Psi}$.
\begin{align}
    \mathbf{E}_0 &= \mathbf{\Psi}^{\mathrm{H}} \mathbf{\Delta_E} \mathbf{\Psi}, \quad \mathbf{\Delta_E} = \mathbf{\Delta}_0 \mathbf{D}_{\mathrm{M}} \nonumber \\
    \mathbf{F}_0 & = \mathbf{\Psi}^{\mathrm{H}} \mathbf{\Delta_F} \mathbf{\Psi}, \quad \mathbf{\Delta_F} = \mathbf{\Delta}_0 \sum_{a = 1}^A v_a \mathbf{D}_{\mathrm{I},a} \nonumber \\
    \mathbf{T}_0 &= \mathbf{\Psi}^{\mathrm{H}} \mathbf{\Delta_T} \mathbf{\Psi}, \quad \mathbf{\Delta_T} =\mathbf{\Delta}_0 \sum_{a = 1}^A v_a  \mathbf{D}_{a,0}.
\end{align}
The received signal in \eqref{Gy} can equivalently be re-written as follows (proof in Appendix \ref{Gyreformulated}),
\begin{equation}\label{Gy1}
     \mathbf{G}_0 \mathbf{y}_0 = \sqrt{p_0} \mathbf{E}_0 \mathbf{s}_0 + \mathbf{E}_0 \sum_{i = 1}^{U} \sqrt{p_i} \mathbf{s}_i + \mathbf{F}_0 \mathbf{s} +\mathbf{z}_0.
\end{equation}
Therefore, the SNR for detecting the HM user's signal is expressed as follows,
\begin{equation}\label{snr0}
    \gamma_0 = \frac{p_0 \rho_{\mathrm{T}} \omega_{\mathrm{E}}}{(1-p_0) \rho_{\mathrm{T}} \omega_{\mathrm{E}}  + \rho_{\mathrm{T}} \omega_{\mathrm{F}} +  \omega_0 },
\end{equation}
where $\omega_{\mathrm{E}}$, $\omega_{\mathrm{F}}$ and $\omega_0$ are expressed as follows,
% \begin{equation}
%     \omega_{\mathrm{E}}  = \sum_{i = 1}^{NM} \frac{\left| \delta_{{\mathrm{E}},i} \right|^2}{NM},    \quad
%       \omega_{\mathrm{F}}  = \sum_{i = 1}^{NM} \frac{\left| \delta_{{\mathrm{F}},i} \right|^2}{NM},    \quad
%     \omega_0  = \sum_{i = 1}^{NM} \frac{\left| \delta_{0,i} \right|^2}{NM}   ,
% \end{equation}
\begin{align}
     \omega_{\mathrm{E}}  &= \sum_{i = 1}^{NM} \frac{\left| \delta_{{\mathrm{E}},i} \right|^2}{NM},  \nonumber  \\
      \omega_{\mathrm{F}}  &= \sum_{i = 1}^{NM} \frac{\left| \delta_{{\mathrm{F}},i} \right|^2}{NM},     \nonumber  \\
    \omega_0  &= \sum_{i = 1}^{NM} \frac{\left| \delta_{0,i} \right|^2}{NM}   ,
\end{align}
where $\delta_{{\mathrm{E},i}}$ and $\delta_{{\mathrm{F},i}}$ are the elements of $\mathbf{\Delta_{E}}$ and $\mathbf{\Delta_{F}}$, respectively expressed as follows,
\begin{align}
     \delta_{{\mathrm{E},i}} &= \delta_{0,i} \sum_{a=1}^A v_a\lambda_{a,i}^M , \nonumber \\
    \delta_{{\mathrm{F},i}} &= \delta_{0,i} \sum_{a=1}^A v_a d_{a,i},
\end{align}
where $d_{a,i}$ are the elements of $\mathbf{D}_{{\mathrm{I}},a}$.
%(See Appendix \ref{snr00} for the proof).
From the SNR expression, we can observe that increasing the OTFS resolution improves the detection SNR at the expense of higher dimensions of the channel matrices, thus, higher computational complexity.

\subsection{Signals Detection at LM users}
In order to detect the HM user at the LM user's side, we consider the following detection matrix,
\begin{equation}
    \mathbf{G}_u = \left( {\mathbf{H}}_u^\mathrm{H} \mathbf{H}_u + \rho \mathbf{I}_{NM}  \right)^{-1} {\mathbf{H}}_u^\mathrm{H}.
\end{equation}
If we consider the diagonalization of $\mathbf{H}_u = \mathbf{\Psi}^\mathrm{H} \mathbf{D}_u \mathbf{\Psi}$, $\mathbf{D}_u$ comprising the eigenvalues of $\mathbf{H}_u$, the detection matrix can be equivalently written as follows,
\begin{equation}
    \mathbf{G}_u = \mathbf{\Psi}^{\mathrm{H}} \mathbf{\Delta}_u \mathbf{\Psi},
\end{equation}
where $\mathbf{\Delta}_u$ is a diagonal matrix consisting of elements,
\begin{equation}
    \delta_{u,i} = \frac{\sum_{a = 1}^A   ({v_a} \lambda_{a,i}^u)^*}{ \left |\sum_{a = 1}^A  v_{a} \lambda_{a,i}^u \right|^2 + \rho }, 1 \le i \le NM,
\end{equation}
$\lambda_{i,a}^u$ are the eigenvalues of $\mathbf{H}_{u,a}$.
Therefore, after applying the equalizer to the received signal in \eqref{yu}, we get the following signal,
\begin{equation}
    \mathbf{G}_u y_u = \sqrt{p_0} \mathbf{G}_u \mathbf{H}_u \mathbf{s}_0 + \mathbf{G}_u \mathbf{H}_u \sum_{i = 1}^{U}  \sqrt{p_i} \mathbf{s}_i + \mathbf{G}_u \mathbf{w}_u.
\end{equation}
Using the same steps to derive \eqref{snr0} and assuming perfect detection, the detection SNR of the HM user at the LM user is expressed as follows,
\begin{equation}
    \gamma_{0,u} = \frac{p_0 \rho_{\mathrm{T}} \omega_{T}}{ \sum_{i = 1}^{U} p_i \rho_{\mathrm{T}}  \omega_{T} + \omega_\mathrm{U}}, 
\end{equation}
where $\omega_T$ and $\omega_U$ can be written as,
\begin{align}
    \omega_\mathrm{T} &=  \sum_{i = 1}^{NM} \frac{\left | \delta_{u,i} \right|^2}{NM} \left|\sum_{a = 1}^A   {v_a} \lambda_{a,i}^u \right|^2, \nonumber \\
    \omega_\mathrm{U} &=  \sum_{i = 1}^{NM} \frac{\left | \delta_{u,i} \right|^2}{NM} .
\end{align}
The LM user successfully decodes and substracts the HM user's signal. At instant $nT$ and subchannel $m \Delta f$, the received signal at each LM user $u$ in the TF domains is expressed as follows,
%each LM user $u$ observes the following signal in the TF domain,
\begin{align}\label{yu1}
    y_u[n,m] &= \sum_{a = 1}^A v_{a} {H}_{u,a}[n,m] \sum_{i = 1}^{U}  \sqrt{p_i} s_{i}[n,m] + w_u[n,m],
\end{align}
where the channel gain and noise are expressed in the TF domain. The channel gain of the LM user is time-independent (not subject to Doppler effect) and LM users access orthogonal subchannels.
Each user $u$ accesses the received signal on its allocated subchannel as in \eqref{su}. %The LM users access orthogonal subchannels 
%Since LM users signals in the same picocell are orthogonal in the frequency domain as in \eqref{suq}, user $u$ considers only the received signal on its dedicated suchannel $m$ and disregards the other suchannels. 
We use a single-tap equalizer to detect the LM user's signal since we assume its channel is flat fading over its dedicated subchannel, as follows,
%Moreover, we assume that the coherence bandwidth of the LM user's channel is larger than the bandwidth of each suchannel meaning that the signal experiences a flat fading channel.%\footnote{Herein, we use ideal pulse shaping resulting in the absence of ICI.}. 
%As a result, we can use a single-tap equalizer to detect the LM user's signal as follows,
\begin{align}
    \frac{y_u[n,u-1]}{H_u[u-1]} = \sqrt{p_u} s_u[n,u-1] 
   + \frac{w_u[n,u-1]}{H_u[u-1]},
\end{align}
where $H_u[u-1] = \sum_{a = 1}^A v_a H_{a,u}[u-1]$ and $H_{a,u}$ is expressed as,
\begin{equation}
    H_{a,u}[m] = \sum_{p=1}^{L_u} \alpha_{p,u} e^{j2\pi \frac{l_{{p},u} m}{M}},
\end{equation}
where $l_{p,u}$ is the delay tap of path $p$ expressed as in \eqref{taps}. 
 Each LM user detects its signal with the following SINR,
\begin{equation}\label{gammau}
    \gamma_u = p_u \rho_{\mathrm{T}} \left| H_u[u-1] \right|^2.
\end{equation}

  \begin{figure}[!h]
	\centering 
	\captionsetup{justification=centering}
	\includegraphics[width=3.5in]{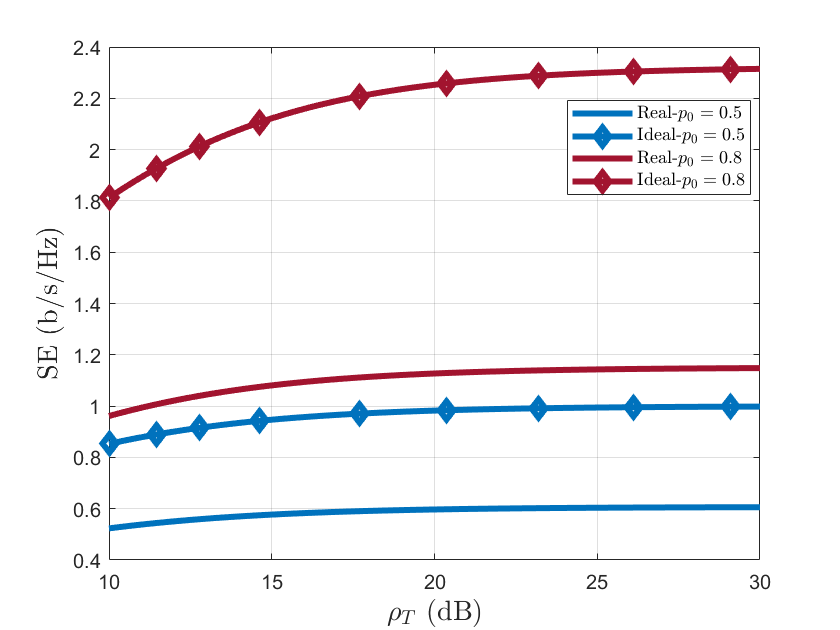}
	\caption{Performance at HM side.}
	\label{hm}
\end{figure}

\section{Simulation Results}
In this section, we present the simulation results for the proposed OTFS-NOMA system. We use the following parameters unless otherwise specified:
%In this section, we evaluate the performance of the proposed system considering the following parameters unless otherwise specified: 
 $A = 4, N = 16, M = 16, \Delta f = 15 \; \mathrm{kHz}, f_{\mathrm{c}} = 5 \; \mathrm{GHz}$, $\rho = 1$, $N_0 = 5$. The number of paths of the HM user's channel is $L_0 = 5$. The maximum Doppler shift is 2.314 KHz corresponding to a maximum speed of 500 km/h.
 % The Doppler shifts in Hz of each path are: $\,[ 2300 ,183, 2288, 2003, 1322 \,]$.
 %\cite{shi2023integrated}. 
% We generate Doppler taps following Jake's model \cite{raviteja2018interference} with 500 km/h a maximum HM speed and the angles randomly selected between $1^o$ and $90^o$.
The maximum channel delay tap is $l_{\max} = 4$. The number of paths of LM users' channels are randomly selected between 1 and 4. 
%The number of paths is randomly generated uniformly between 1 and 5. 
%The number of LM users is uniformly distributed between 1 and $U_{\max}$, where $U_{\max} = 8$.
The number of LM users is $U = 8$.
The LM user with better channel condition is allocated less transmit power. Thus, $p_u = (1-p_0) \frac{1/|H_u|}{\sum_{i = 1}^U 1/|H_i|}, \forall 1 \le u \le U$.
The outage probability is computed for a minimum rate threshold of $R_\mathrm{th} = 0.5 \; \mathrm{b/s/Hz}$.

We first evaluate the impact of the IDI caused by fractional Doppler on the performance of the HM user. To this end, we plot in Fig. \ref{hm}, the spectral efficiency of the HM user for two different values of NOMA allocation factors: $p_0 = 0.5$ and $p_0 = 0.8$. In the figure, ``Real" and ``Ideal" refer to assuming the presence and absence of fractional Doppler, respectively. As expected, the HM user's spectral efficiency degrades when the IDI is considered as there is an additional interference term in the SNR expression. Therefore, neglecting the effect of fractional Doppler on OTFS modulated signals gives misleading information on the system performance, which may be detrimental in terms of unaccounted-for outage events. Moreover, we notice that when $p_0 = 0.5$, the difference in spectral efficiency between the IDI-absence and IDI-presence cases is around 0.4 b/s/Hz. On the other hand, when $p_0 = 0.8$, the difference increases to 1 b/s/Hz. Interestingly, this difference does not vary with the transmit SNR.
Therefore, the gap between the real and ideal performances becomes more noticeable when the HM user NOMA power factor increases. Thus, when the IDI is not considered, although increasing the HM user's transmit power improves its spectral efficiency, it results in increased errors and unexpected deteriorating performance.
%Furthermore, the detection spectral efficiency of the HM user at the HM user side converges to a constant when the transmit SNR increases. This behavior explains the fact that the performance gap between the ``Real" and ``Ideal" cases is almost a constant as function of $\rho_{\mathrm{T}}$.

  \begin{figure}[!h]
	\centering 
	\captionsetup{justification=centering,margin=1cm}
	\includegraphics[width=3.5in]{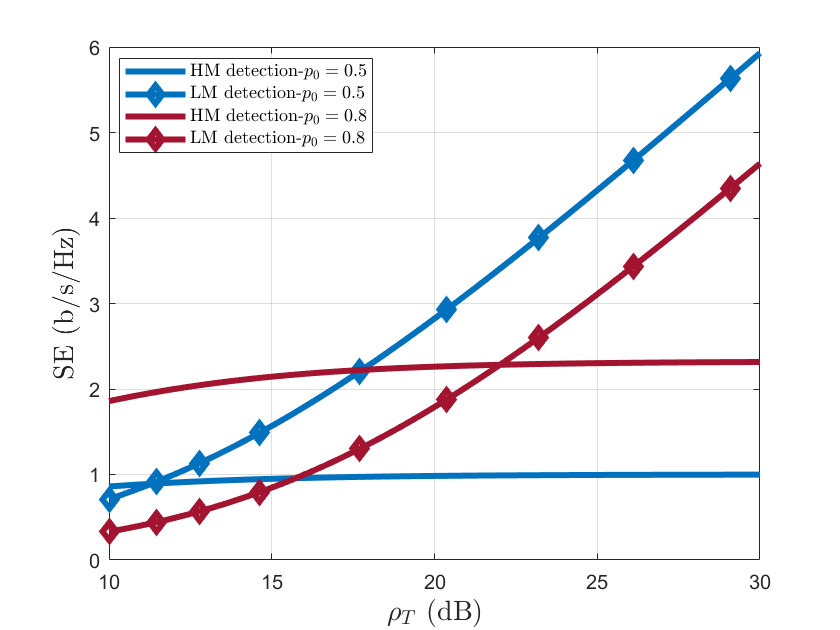}
	\caption{Performance at LM side.}
	\label{lm}
\end{figure}
Next, we study the worst-case LM users performance by considering the average minimum performance at the LM user's side. We plot in Fig. \ref{lm}, the spectral efficiency of the HM user's signal detection and LM user's signal detection for two different NOMA power factors: $p_0 = 0.5$ and $p_0 = 0.8$. The HM detection occurs at the LM side with a significantly lower SNR than the LM signal's detection because we assume that the successive interference cancellation is conducted without errors. Moreover, when the LM user is allocated a higher transmit power, it achieves greater spectral efficiency gain, which is constant across all values of the total SNR $\rho_{\mathrm{T}}$. This gain is around 1 b/s/Hz for both HM and LM signals detections since all LM users share the remaining power, contrarily to the HM user's performance in the previous example, where the HM power boost resulted in a significant quality-of-service improvement. Therefore, this shows that it is essential to strategically tune the NOMA power factors according to different user's requirements. 

  \begin{figure}[!h]
	\centering 
	\captionsetup{justification=centering,margin=1cm}
	\includegraphics[width=3.5in]{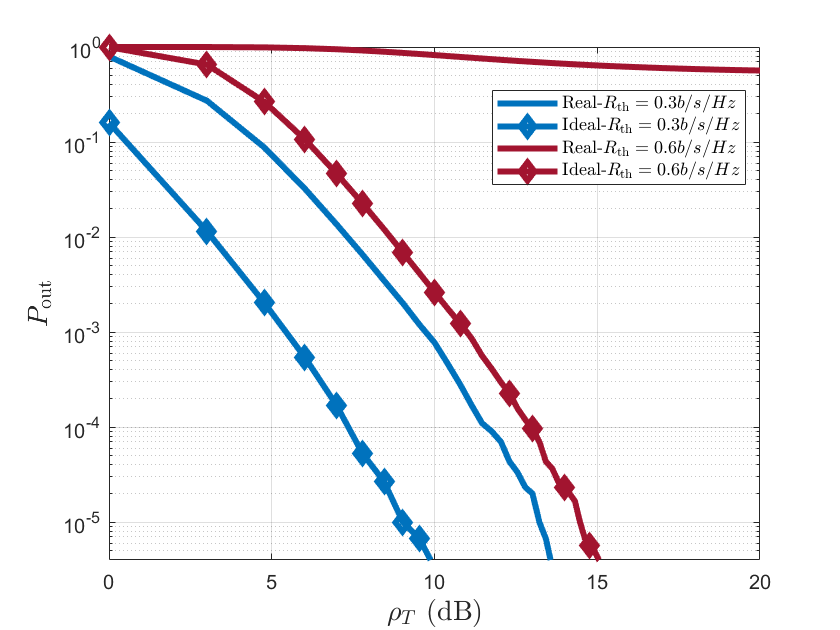}
	\caption{Outage probability results.}
	\label{out}
\end{figure}

Finally, we evaluate the performance of the OTFS-NOMA system in terms of outage events. In Fig. \ref{out}, we plot the outage probability of the HM user for two different values of minimum rate requirement $R_{\mathrm{th}} = 0.3 \; b/s/Hz$ and $R_{\mathrm{th}} = 0.6 \; b/s/Hz$. These results are evaluated for both assumptions of presence and absence of IDI and for $p_0 = 0.5$.
First, we notice that the HM user's performance difference between the ``Real" and ``Ideal" cases is significant in terms of outage probabilities. For instance, when  $\rho_{\mathrm{T}} = 10 \; \mathrm{dB}$ and $R_{\mathrm{th}} = 0.3 \; b/s/Hz$, the outage probability when IDI is taken into account is 3e-6 while it is in the order of 8e-4 when the IDI is not taken into account. As a result, if IDI is present and not taken into account, the system may seem to perform much better than it does in practice.  
Furthermore, the performance gap between the ``Real" and ``Ideal" cases behaves in different ways when the transmit SNR increases. In fact, this gap significantly increases with $\rho_{\mathrm{T}}$ when $R_{\mathrm{th}} = 0.6 \; b/s/Hz$, differently from the case of $R_{\mathrm{th}} = 0.3 \; b/s/Hz$. 
This can be explained by the fact that the average spectral efficiency is above the threshold 0.3 b/s/Hz for both ``Real" and ``Ideal" cases. On the other hand, the average spectral efficiency is below the threshold 0.6 b/s/Hz in the ``Real" case and significantly above this threshold in the ``Ideal" case (See Fig. \ref{hm}).

% In the last simulation example, we investigate the effect of the number of subpaths generated from each path on the HM user's performance. In Fig. \ref{np}, we plot the spectral efficiency of the HM user for two different values of power alloction $p_0$ and for $N_0 = 1$ and $N_0 = 10$. First, we notice that the HM user's communication link achieves less spectral efficiency when the number of subpaths is higher because of the increased IDI interference. Second, we notice that the performance gap between the two values of interfering subpaths is higher when the HM user's NOMA power factor is higher. is allocated more transmit power. This can be explained by the fact that allocating more transmit power to the HM user will also increase the cumulative power of the interfering signals and therefore increase the impact of the IDI on the received SNR. As a result, if the number of subpaths is not known at the transmitter, decreasing the HM allocated power will make the effect of this parameter on the system performance less significant. In this case, w specific number can be chosen. Certainly, this measure comes at the expense of decreased HM performance in terms of spectral efficiency. 

\section{Conclusion}
In this work, we investigated the impact of fractional Doppler on the performance of the OTFS-NOMA system. The resulting IDI proved to have a significant impact on the performance of the HM user in terms of spectral efficiency, especially for higher HM user power allocation factors. Moreover, outage events can deteriorate the performance of the HM user if fractional Doppler is not considered during system modeling. 
Therefore, it is important to take this additional source of interference into account in order to make successful decisions during resource distribution between different users.
In the presence of fractional Doppler, NOMA paradigm plays an important role in the considered multi-user system in many ways. First, it helps accomodate multiple users with different mobility profiles into the same resource block. Second, it provides a convenient degree of freedom to accommodate different users' requirements in the presence of IDI. 
\begin{appendices}
\section{Reformulated Equalized Signal} 
\label{Gyreformulated}
Since $\mathbf{D}_{a,0} = \mathbf{D}_{\mathrm{M},a} +  \mathbf{D}_{\mathrm{I},a}$, the matrix $\mathbf{T_0}$ can be reformulated as,
\begin{align}
    \mathbf{T_0} &= \mathbf{\Psi}^{\mathrm{H}} \mathbf{\Delta}_0 \sum_{a = 1}^A v_a \mathbf{D}_{\mathrm{M},a} \mathbf{\Psi} + \mathbf{\Psi}^{\mathrm{H}} \mathbf{\Delta}_0 \sum_{a = 1}^A v_a \mathbf{D}_{\mathrm{I},a} \mathbf{\Psi} \nonumber \\
    &= \mathbf{\Psi}^{\mathrm{H}} \mathbf{\Delta}_0 \mathbf{D}_{\mathrm{M}} \mathbf{\Psi} + \mathbf{F_0} 
     = \mathbf{E_0} + \mathbf{F_0}.
\end{align}
Since $\mathbf{s} =  \sum_{i = 0}^{U} \sqrt{p_i} \mathbf{s}_i$, we get the expression in \eqref{Gy1}.

\end{appendices}

\bibliographystyle{IEEEtran}
\bibliography{IEEEabrv,references}
\end{document}